\documentclass[nonacm,sigplan]{acmart}

\usepackage[]{hyperref}
\makeatletter
\newif\if@restonecol
\makeatother

\usepackage[linesnumbered,ruled,vlined]{algorithm2e}

\usepackage[normalem]{ulem}
\usepackage{amsmath}
\usepackage{enumitem}

\usepackage{listings}
\usepackage{xcolor}

\definecolor{codegreen}{rgb}{0,0.6,0}
\definecolor{codegray}{rgb}{0.5,0.5,0.5}
\definecolor{codepurple}{rgb}{0.58,0,0.82}
\definecolor{backcolour}{rgb}{0.95,0.95,0.92}

\lstdefinestyle{mystyle}{
    commentstyle=\color{codegreen},
    keywordstyle=\color{magenta},
    numberstyle=\tiny\color{codegray},
    stringstyle=\color{codepurple},
    basicstyle=\footnotesize,
    breakatwhitespace=false,         
    breaklines=true,                 
    captionpos=b,                    
    keepspaces=true,                 
    numbers=left,                    
    numbersep=5pt,                  
    showspaces=false,                
    showstringspaces=false,
    showtabs=false,                  
    tabsize=2
}

\begin{document}
\pagestyle{empty}
\title{FaaSTube: Optimizing GPU-oriented Data Transfer for Serverless Computing}


\begin{abstract}

Serverless computing has gained significant traction for machine learning inference applications, which are often deployed as serverless workflows consisting of multiple CPU and GPU functions with data dependency.
However, existing data-passing solutions for serverless computing primarily reply on host memory for fast data transfer, mandating substantial data movement and resulting in salient I/O overhead.
In this paper, we present FaaSTube, a GPU-efficient data passing system for serverless inference.
FaaSTube manages intermediate data within a GPU memory pool to facilitate direct data exchange between GPU functions. It enables fine-grained bandwidth sharing over PCIe and NVLink, minimizing data-passing latency for both host-to-GPU and GPU-to-GPU while providing performance isolation between functions. 
Additionally, FaaSTube implements an elastic GPU memory pool that dynamically scales to accommodate varying data-passing demands. Evaluations on real-world applications show that FaaSTube reduces end-to-end latency by up to 90\% and achieves up to 12x higher throughput compared to the state-of-the-art.

\end{abstract}

\author{{\rm Hao Wu$^{\dagger}$, Junxiao Deng$^{\dagger}$, Minchen Yu$^{\S}$, Yue Yu$^{\dagger}$, Yaochen Liu$^{\dagger}$, Hao Fan$^{\dagger}$, Song Wu$^{\dagger}$, Wei Wang$^{\ddagger}$}\\
   {\em $^{\dagger}$National Engineering Research Center for Big Data Technology and System,\\ Services Computing Technology and System Lab, Cluster and Grid Computing Lab,\\ School of Computer Science and Technology, Huazhong University of Science and Technology, China}\\
         {\em $^{\ddagger}$Hong Kong University of Science and Technology}\\ 
    	{\em $^{\S}$The Chinese University of Hong Kong, Shenzhen}
     \vspace{0.1cm}
    }

\maketitle

\section{Introduction}\label{introduction}

The widespread adoption of Machine Learning (ML) inference applications has heightened the demand for efficient inference service systems~\cite{Reef,deepplan,GPUlet,SHEPHERD}. Recent research suggests deploying inference applications on GPU-enabled serverless platforms~\cite{infless,Batch,llama,tetris,streambox,faaswap}. With serverless computing, users package ML models as functions and leave resource provisioning and scaling to the serverless platform. This approach enables users to concentrate on the development of application logic while maintaining resource efficiency, eliminating the need for overprovisioning.

In real-world scenarios, inference applications often stitch together multiple models and operations into a workflow. Table 1 presents several representative inference applications from recent research~\cite{Boggard,cocktail,Adainf,infline,scrooge}. As a concrete example, in a traffic monitoring application~\cite{Adainf} that analyzes pedestrian and vehicle traffic (Fig.~\ref{workflows}), video frames are first decoded and preprocessed. A detection model then extracts objects from these frames. The sub-images of detected pedestrians and vehicles are subsequently passed to two separate recognition models for further analysis of behavior and type. Inference applications in serverless computing are structured as \emph{serverless inference workflows}. These workflows are hybrid, consisting of GPU functions (gFunc), CPU functions (cFunc), and their data dependencies.

\begin{figure}[t]
\centerline{\includegraphics[width=0.48\textwidth]{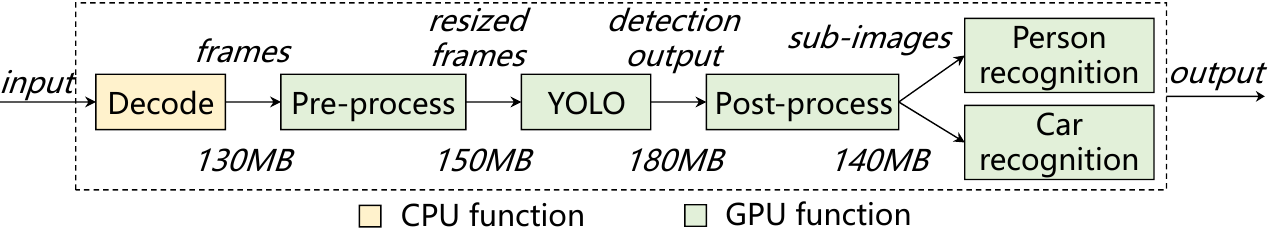}}
\caption{A typical traffic analysis application~\cite{Adainf,Boggard}.}
\label{traffic workflow}
\end{figure}

\begin{figure}[t]
\centerline{\includegraphics[width=0.45\textwidth]{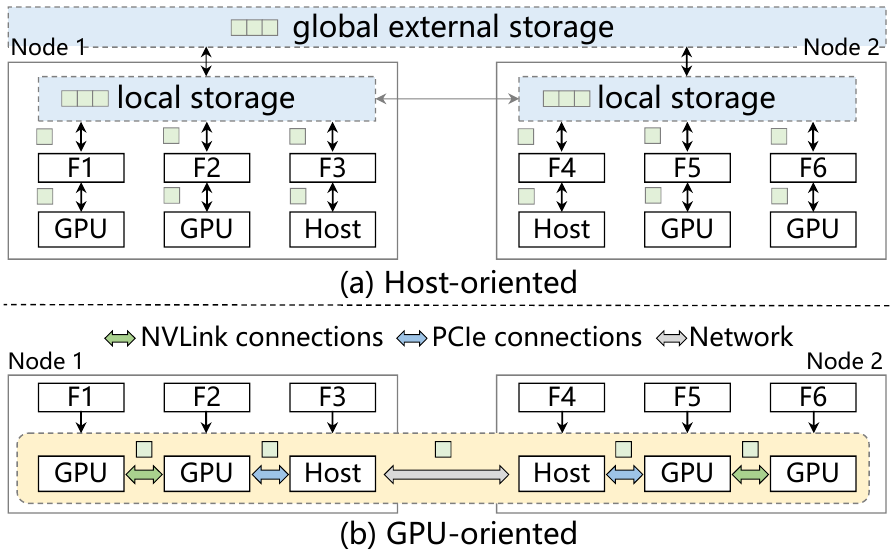}}
\caption{Comparison of host-oriented inter-function data passing and our GPU-oriented inter-function data passing.}
\label{compare}
\end{figure}

Serverless inference workflows involve various types of data passing. In addition to typical cFunc-to-cFunc data passing, there are host-to-gFunc (where the host represents either a cFunc or I/O data in host memory) and gFunc-to-gFunc data passing. Unfortunately, current serverless systems rely on a \textbf{Host-oriented} data passing approach (Fig.~\ref{compare}(a)), where intermediate data is stored and exchanged via host memory (i.e., host-side external storage~\cite{dataflower,sonic,nightcore,pheromone,Fuyao}). This host-oriented method overlooks the decoupling of GPU and host memory and the potential of various connections in GPU servers, resulting in substantial unnecessary overhead. For instance, in gFunc-to-gFunc data passing, this method ignores direct NVLinks between GPUs, requiring multiple sequential copies: data is first copied to host memory and then back to the target GPU, leading to large overhead (92\% of the total latency in our experiments). Similarly, for host-to-gFunc data passing, this method only uses a single PCIe link, neglecting the parallel PCIe links available within GPU servers. For instance, parts of the data can be routed to neighboring GPUs via NVLink and then transferred in parallel to host memory through the neighboring GPUs' PCIe links.

Our key idea is to introduce a \textbf{GPU-oriented} data-passing approach for serverless inference workflows. This approach allows data to be stored and exchanged directly in GPU memory, avoiding redundant transfers between GPU and host memory. It also leverages various connections within GPU servers (e.g., parallel PCIe links and NVLinks) to accelerate inter-function data passing, specifically focusing on host-to-gFunc and gFunc-to-gFunc data passing in this paper. To design an efficient GPU-oriented data-passing framework for serverless functions, the following requirements must be met: (1) \emph{Bandwidth efficiency}: The available bandwidth of the various connections must be fully utilized while avoiding contention among concurrent functions. (2) \emph{Topology awareness}: There are different connection topologies in modern GPU servers, requiring careful management of transfer scheduling based on the specific GPU topology during inter-function data passing. (3) \emph{GPU memory efficiency}: As highlighted by existing research~\cite{deepum,HUVM}, GPU memory is a limited resource; therefore, its usage should be minimized when providing GPU data storage for functions. (4) \emph{Transparent deployment}: All of the above should be achieved transparently to users, simplifying the development of serverless inference workflow and hiding the complexities of various connections and GPU topologies involved.


Based on these requirements, we propose FaaSTube, a GPU-efficient data transfer framework designed for serverless inference workflows. FaaSTube acts as a transparent "tube" across GPUs, facilitating efficient data storage and transfer for functions. It comprises three key components:

First, to simplify the management of various data passing in serverless inference workflows, we provide a unified data-passing interface and data index. This allows users to focus on application logic while offloading the complexities of data transfer implementation to FaaSTube. It automatically locates intermediate data within GPU servers and selects the appropriate connections (PCIe, NVLink, or network) and transfer methods (e.g., parallel or pipelined) to deliver data to the requesting function.

Second, in order to fully utilize connections such as PCIe and NVLink for serverless functions, we design effective transfer scheduling mechanisms. (1) To achieve interference-free sharing of PCIe connections among concurrent functions on a shared GPU server, FaaSTube proposes fine-grained PCIe bandwidth isolation. Native GPU PCIe scheduling is not optimized for parallel transfers, and related works~\cite{deepplan,serverlessLLM} focus on exclusive inference systems while overlooking scenarios where multiple functions share a GPU server. FaaSTube manages data transfers across all PCIe connections, flexibly partitioning bandwidth based on each function's SLO and controlling bandwidth usage during transfers. In addition, FaaSTube employs circular pinned memory buffers to enhance PCIe transfer efficiency while minimizing allocation overhead. (2) To ensure robust inter-function data transfer performance across different GPU topologies, FaaSTube implements topology-aware parallel NVLink transfer scheduling. In non-uniform topologies, many GPUs with limited direct NVLink bandwidth can severely restrict the performance of point-to-point (i.e., gFunc-to-gFunc) transfers in serverless inference workflows. Existing works~\cite{MAPA,SCCL,Blink} typically optimize task placement but still rely on a single direct NVLink path, failing to address this issue comprehensively. Other multi-GPU communication methods~\cite{nccl,TACCL,TCCL} focus on collective communication and often ignore point-to-point transfers, also utilizing only one NVLink path. FaaSTube identifies multiple parallel NVLink paths between bandwidth-constrained GPUs, employing contention-aware path selection to accelerate inter-function data passing.

Third, to efficiently manage GPU memory and intermediate data in GPU data store for serverless functions, FaaSTube proposes an auto-scaling GPU memory pool, as function workloads and the size of intermediate data (e.g., the number of objects in a video frame) can vary dynamically. Existing GPU memory management systems~\cite{GMlake,pytorch,deepum} are designed for long-running, exclusive GPU tasks like ML training. Therefore, they often retain a lot of idle memory blocks and lack flexible reclamation mechanisms, leading to excessive memory usage in serverless environments. FaaSTube tracks data storage requirements in real-time and caches only the necessary memory blocks in GPU data store, dynamically resizing the memory pool. Furthermore, when intermediate data accumulates in GPU memory, FaaSTube implements a smart data migration based on function request queue, migrating data to host memory and adaptively prefetching data back to the GPU, thereby effectively alleviating memory pressure without compromising performance.

We implement FaaSTube on top of INFless~\cite{infless}, a recent serverless inference platform built on OpenFaaS. We evaluate FaaSTube on various real-world inference workflows using Azure cloud traces~\cite{Azure-trace}. The results show that FaaSTube reduces end-to-end latency by up to 90\% and achieves up to 12X higher throughput compared to state-of-the-art serverless systems. In addition, we also present the scalability and effectiveness of FaaSTube on a 4-node cluster and GPU servers with various GPU topologies.



\section{Background}

This section introduces serverless inference workflows and analyzes the problem of current inter-function data passing.

\subsection{Serverless Inference Workflow}
ML inference applications are increasingly prevalent in daily life~\cite{Adainf,Boggard,SHEPHERD,scrooge,cocktail,nexus}. To efficiently deploy inference applications, many studies~\cite{infless,llama,Astraea,streambox} have developed GPU-enabled serverless inference systems, allowing users to publish ML models as functions that scale resources on demand with workload fluctuations. Compared to managing GPU clusters directly (i.e., serverful deployment), serverless inference enables users to focus on application logic while eliminating the need for overprovisioning.


Real-world inference applications often involve multiple ML models and operations working together to perform complex tasks. In this paper, we study several representative inference applications collected from recent research~\cite{AQUATOPE,Adainf,cocktail,Boggard,Astraea,infline}, as shown in Table 1 (detailed in Section~\ref{workflows_list}). In serverless systems, these applications combine GPU functions (gFunc) and CPU functions (cFunc) as serverless inference workflows. Serverless inference workflows can be represented as directed acyclic graphs (DAGs), where each node corresponds to ML models on the GPU or operations on the host, with edges representing dataflow between functions. Similar to serverless workflows in data analysis~\cite{faasflow}, serverless inference workflows exhibit diverse dataflows, including sequence, condition, and fan-in/fan-out patterns.

\begin{table}[t]
\centering
\caption{Real-world inference workflows. GPU functions are underlined, and there are four typical workflow types: \emph{condition}, \emph{sequence}, \emph{fan-in}, and \emph{fan-out}.}
\setlength{\tabcolsep}{0.5pt}
{
\footnotesize
\centering
\begin{tabular}{ c |c |c }
\hline

\textbf{GPU workflows} & \textbf{Functions} & \textbf{Type} \\
\hline

Traffic monitor~\cite{Boggard} &  decode, \uline{pre/postproc.}, \uline{Yolo-det}, \uline{ResNet} &  condition\\
Auto-driving~\cite{Adainf} &  decode, \uline{denoising}, \uline{Yolo-seg}, bluring &  sequence \\
Video editing~\cite{AQUATOPE}  &  decode, \uline{preproc}, \uline{Yolo-face}, \uline{ResNet}  & fan-in \\
Image editing~\cite{cocktail} &  decode, \uline{denoising}, \uline{ResNet}, \uline{AlexNet} &  fan-out \\
Social media~\cite{infline} &  decode, \uline{preprocess}, \uline{OCR}, \uline{Bert} &  condition\\
Yelp~\cite{Astraea} & \uline{Bert}, \uline{Bert} & sequence \\
\hline
\end{tabular}
}\label{workflows}
\end{table}

\subsection{Current Inter-Function Data Passing}
Since media data (e.g., images and video frames) can reach hundreds of MB, efficient data passing is crucial in serverless inference workflows. However, passing data between different types of functions in these workflows is non-trivial. Beyond typical cFunc-to-cFunc data passing, there are also host-to-gFunc (where the host represents a cFunc or I/O data in host memory) and gFunc-to-gFunc data passing. Unfortunately, current serverless systems rely on a \textbf{host-oriented} data passing approach (Fig.~\ref{compare}(a)), where intermediate data is stored and exchanged via an external storage in host memory, such as remote data storage (e.g., AWS S3~\cite{awss3}) or local data storage (e.g., intra-node message queues~\cite{faasflow,sonic,dataflower} or shared memory~\cite{pheromone,nightcore,RDMMap}). This method ignores the decoupling of GPU and host memory, forcing the intermediate data of GPU function to be exchanged through host memory, resulting in large overhead.

\begin{figure}[b]
\centerline{\includegraphics[width=0.45\textwidth]{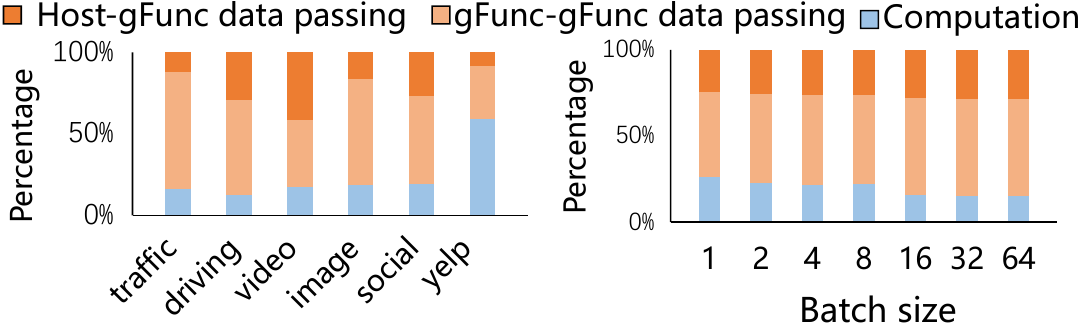}}
\caption{Performance analysis of real-world inference workflows on INFless+. (a) Breaking down of overall latency. (b) Breaking down of latency for \emph{Traffic} workflow with various batch sizes. Each bar is broken into three parts: the latencies of host-to-gFunc data passing (top), gFunc-to-gFunc data passing (middle), and computation (bottom).
}
\label{ob_1}

\end{figure}

Next, we evaluate the latency of INFless+, which combines the latest serverless inference system, INFless~\cite{infless}, with recent data passing optimizations for serverless workflows in Pheromone~\cite{pheromone}, i.e., sharing host memory between functions. Experiments are conducted using a set of real-world inference applications (Table~\ref{workflows}) on a DGX-V100 GPU server. Additional details can be found in Section~\ref{workflows_list}. Fig.~\ref{ob_1} shows that inter-function data passing accounts for up to 92\% of the end-to-end latency, with 29\% of host-to-gFunc data passing, 63\% of gFunc-to-gFunc data passing. This high latency arises from the overhead of copying data between GPU and host memory, as each GPU is connected to host memory via PCIe links with limited bandwidth (12GB/s under PCIe 3.0). The gFunc-to-gFunc transfer overhead is especially high, as it involves two sequential data copies: first from the source GPU to host, then from host to the target GPU. Since functions share host memory, the overhead of cFunc-to-cFunc will be less than 1\%. Therefore, it is omitted from the figure.



\section{GPU-oriented Data Passing}

This section presents the benefits of exploiting various connections (i.e., parallel PCIe links and direct NVLinks) in GPU servers and analyzes the associated challenges.

\subsection{Opportunities and Requirements }

\begin{figure}[b]
\centerline{\includegraphics[width=0.4\textwidth]{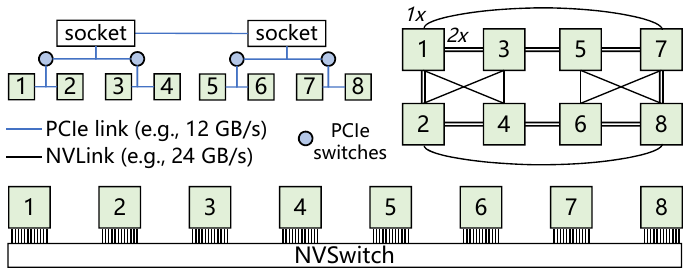}}
\caption{Vairous connection topologies in GPU servers: (a) PCIe connections between host and GPUs, (b) 8 GPUs connected via hard-wired NVLinks like DGX V100 and (c) 8 GPUs connected via switch-based NVLinks like DGX A100.}
\label{topo}

\end{figure}

Modern GPU servers are equipped with high-speed NVLink connections between GPUs and multiple PCIe connections between host and GPUs, facilitating faster data transfer in serverless inference workflows. For instance, in gFunc-to-gFunc data passing, data can be transferred directly via NVLink (25-50 GB/s), avoiding unnecessary copies between the host and GPUs. In host-to-gFunc data passing, part of the data can be routed to a neighboring GPU via NVLink and then transferred to the host in parallel through the neighboring GPU's PCIe link. As shown in Fig~\ref{topo}, DGX-V100 and DGX-A100 servers connect 8 GPUs to the host via 4 PCIe links, allowing parallel transfers with up to 4x bandwidth. 

However, existing host-oriented inter-function data passing overlooks these opportunities. This motivates us to propose a GPU-oriented inter-function data passing approach that allows intermediate data to be exchanged or stored directly in GPU memory, thereby avoiding costly data copies between the host and GPUs. Additionally, we leverage various connections available in GPU servers, such as parallel PCIe links and NVLinks, to accelerate inter-function data passing. Nevertheless, designing an efficient GPU-oriented data passing framework for serverless functions needs to meet several key requirements:

\emph{\textbf{R1}: Bandwidth efficiency}: The available bandwidth of the various connections must be fully utilized while avoiding contention among concurrent functions.

\emph{\textbf{R2}: Topology awareness}: There are different connection topologies in modern GPU servers, requiring careful management of transfer scheduling and link usage based on the specific GPU topology during inter-function data passing.

\emph{\textbf{R3}: GPU memory efficiency}: As highlighted by existing research~\cite{deepum,HUVM}, GPU memory is a limited resource; therefore, its usage should be minimized when designing GPU data storage for functions.

\emph{\textbf{R4}: Transparent deployment}: All of the above should be implemented transparently to users, simplifying the development of serverless workflow and hiding the complexities of diverse connections and GPU topologies.

\subsection{Technical Challenges}\label{challenges}
Fulfilling the above requirements still faces several challenges that existing methods fail to address.

\noindent\textbf{Challenge \#1: PCIe connections sharing}. \emph{Harvesting parallel PCIe links necessitates avoiding bandwidth contention among concurrent functions and implementing effective pinned memory allocation (\textbf{R1}).}

\begin{figure}[b]
\centerline{\includegraphics[width=0.45\textwidth]{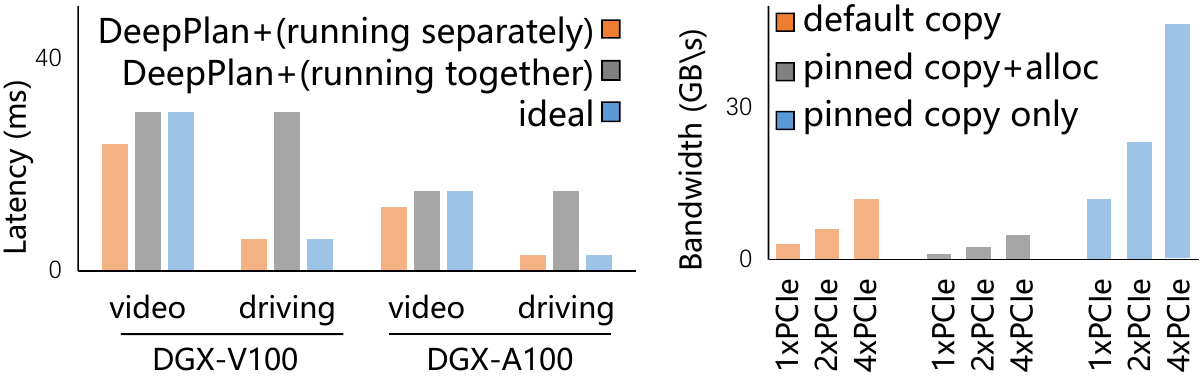}}
\caption{(a) Comparison of host-to-gFunc passing overhead between separate execution and together execution of video and driving workflows. (b) The Impact of pinned memory in PCIe transfers.}
\label{ob_2}

\end{figure}

When leveraging parallel PCIe links for host-to-gFunc data passing, current PCIe transfer scheduling faces two limitations. First, borrowing multiple PCIe links introduces \emph{interference} when concurrent functions on neighboring GPUs simultaneously transfer data to the host. However, related works~\cite{deepplan,serverlessLLM} that utilize parallel PCIe links focus on exclusive inference systems and overlook scenarios where multiple functions share a GPU server. Additionally, native GPU PCIe scheduling is not optimized for parallel transfers. We evaluated the performance of concurrent workflows in DeepPlan+, which implements parallel PCIe transfers as in DeepPlan~\cite{deepplan} on INFless+. Fig.~\ref{ob_2}(a) shows that running the video and driving workflows concurrently leads to a 5x increase in host-to-gFunc transfer overhead for the driving workflow compared to running them separately. This increase is due to the video workflow's multiple functions loading video blocks simultaneously, saturating the global PCIe bandwidth. Such interference can significantly impact the performance of SLO-critical functions. Second, \emph{pinned memory} is needed to maximize PCIe link utilization, as it increases each link's bandwidth from 3 GB/s to 12 GB/s. However, pinned memory allocation incurs large overhead (e.g., 70 ms for 100 MB), reducing transfer bandwidth to 1 GB/s, as shown in Figure ~\ref{ob_2}(b). This overhead greatly impacts the performance of short-lived functions.

\noindent\textbf{Challenge \#2: Various topologies of GPUs and NVLinks}. \emph{Function workflows rely on point-to-point transfers, making them more sensitive to topology and NVLink usage (\textbf{R1}, \textbf{R2}).}

\begin{figure}[t]
\centerline{\includegraphics[width=0.45\textwidth]{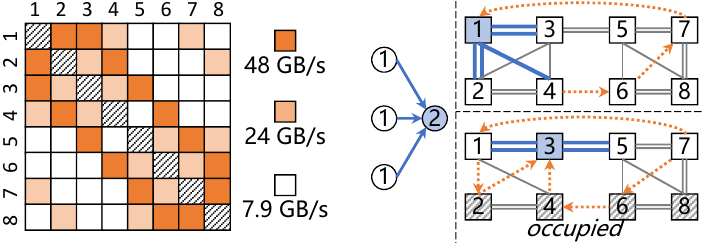}}
\caption{The non-uniform GPU topology. (a) The P2P bandwidth of different GPU pairs in a DGX-V100 server. (b) Different placement of a video workflow. Blue lines show the currently used paths, and orange dotted lines show possible parallel paths.}
\label{ob_3}

\end{figure}

Modern GPU server have diverse topologies with NVLinks. In high-end GPU servers, GPUs are fully connected via NVSwitch (Fig~\ref{topo}(c)). In contrast, more cost-effective GPU servers commonly exhibit non-uniform topologies with hard-wired NVLinks (Fig~\ref{topo}(b)). Recent production traces~\cite{mlaas,gandiva} indicate that this kind of GPU servers comprise 36\% of the cluster, and many cloud providers offer them as rental options (AWS P3 instance~\cite{awsp3}, Azure NDv2 instance~\cite{Azurecloud}, Google cloud N1 instance~\cite{Googlecloud}). In non-uniform topologies, point-to-point transfer bandwidth between GPUs can vary significantly. As shown in Fig.~\ref{ob_3}(a), 28\% of GPU pairs have only half the bandwidth, while 42\% lack a direct NVLink and must transfer data over PCIe with a bandwidth of just 7.9 GB/s.

When functions in a serverless workflow are deployed on GPUs with limited point-to-point bandwidth, performance can significantly degrade. Existing methods~\cite{Blink,SCCL,MAPA} optimize placement to avoid transfers between these GPUs. However, in complex workflows, e.g., traffic(condition), video (fan-in), and image (fan-out), using bandwidth-limited GPUs is unavoidable due to connection limitations. Consider a video processing workflow that includes three parallel detection functions for frame extraction and an identification function for actors. Even with optimal deployment (top of Fig.~\ref{ob_3}(b)), this workflow still involves bandwidth-limited GPU pairs, such as $G_1$ and $G_4$. In a worst-case (bottom of Fig.~\ref{ob_3}(b)), optimal deployment may be impossible (e.g., $G_2$, $G_4$, $G_6$ and $G_8$ are occupied by other workflows), forcing $G_7$ and $G_3$ to participate with no NVLink.

This problem cannot be completely solved by placement optimisation alone, as it relies on only the direct NVLink path between GPUs. We propose viewing a GPU server as a network where each GPU pair has parallel NVLink paths. For instance, routing G1-G4 through G4-G6-G7 can double the bandwidth. In addition, G3-G7 can utilize paths like G7-G1-G2-G3 and G7-G6-G4-G3, increasing bandwidth by 6X. Existing multi-GPU communication methods~\cite{TACCL,TCCL}, such as NCCL~\cite{nccl}, focus on collective communication and also utilize only a single NVLink path for point-to-point transfers.

\noindent\textbf{Challenge \#3: Efficient GPU data store}. \emph{Fluctuations in function workload and intermediate data size require an elastic GPU data store that can dynamically scale and efficiently migrate data between GPUs and host memory (\textbf{R3}).}

\begin{figure}[t]
\centerline{\includegraphics[width=0.47\textwidth]{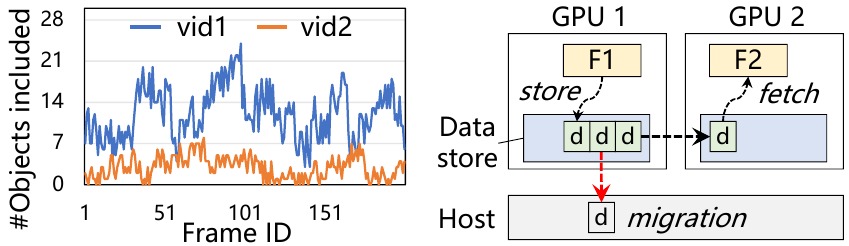}}
\caption{(a) The object count fluctuates in different videos. (b) The accumulation of intermediate data results in its migration to host memory.}
\label{ob_4}

\end{figure}

To avoid storing data in host memory, a GPU data store is necessary. As many studies highlight~\cite{GMlake,deepum,TGS}, GPU memory is limited, so this data store must be space-efficient. However, managing the GPU data store for serverless workflows needs to adapt to two types of fluctuations: 1) \emph{Data size fluctuations}: In media-based workflows, the size of intermediate data varies with the semantic content of the input. For example, in a traffic monitoring workflow, the detection function extracts objects from each frame and passes them to the recognition function. As shown in Fig.~\ref{ob_4}(a), the number of objects in different video frames fluctuates with pedestrian and vehicle movements. Moreover, the intermediate data size varies with different input batch sizes (e.g., TensorRT dynamic batching~\cite{dynamicshape}). Therefore, the GPU data store must allocate memory flexibly and dynamically scale based on function's demand. 2) \emph{Memory pressure fluctuations}: Large amounts of cached intermediate data can occupy a lot of GPU memory, particularly when production rates exceed consumption, leading to data accumulation, or when functions face bursty requests. In such cases, the GPU data store must migrate data to host memory as necessary (Fig.~\ref{ob_4}(b)).

However, existing GPU memory management, such as memory pooling~\cite{pytorch,GMlake} and unified memory~\cite{HUVM, Unifiedmem}, are designed for long-running, exclusive tasks (e.g., ML training). Therefore, they often retain significant idle memory blocks and lack the necessary elasticity for serverless functions, particularly in terms of flexible memory recycling mechanisms (Section~\ref{elastic_mempool}), resulting in excessive memory occupation.

\noindent\textbf{Challenge \#4: Developer-friendly}. \emph{The development of serverless inference workflows is complicated by the diverse distribution of functions and diverse data passing, involving various connections and memory types in GPU servers (\textbf{R4}).} 

Developing serverless inference workflows presents two challenges. First, \emph{various data storage}: Data may be stored in either host or GPU memory, each requiring distinct management and retrieval mechanisms. Host-side storage typically uses shared memory addresses as Pheromone~\cite{pheromone} or message queue keys (e.g., Redis), whereas GPU-side storage relies on CUDA IPC handles. Second, \emph{various transfer methods}: Serverless functions are distributed across nodes and GPUs, necessitating multiple data transfer methods—such as intra-GPU, inter-GPU, host-GPU, and inter-node—utilizing PCIe, NVLink, or network connections. In addition, optimizations such as parallel and pipelined data transfers need to be selectively implemented to enhance transfer efficiency. What is worse,  serverless functions typically run in containers~\cite{Nvidiacontainer}, which obscures critical information (e.g., GPU topology and function placement), preventing developers from configuring efficient data passing in GPU servers.

\section{Overview of FaaSTube}
\begin{figure}[t]
\centerline{\includegraphics[width=0.48\textwidth]{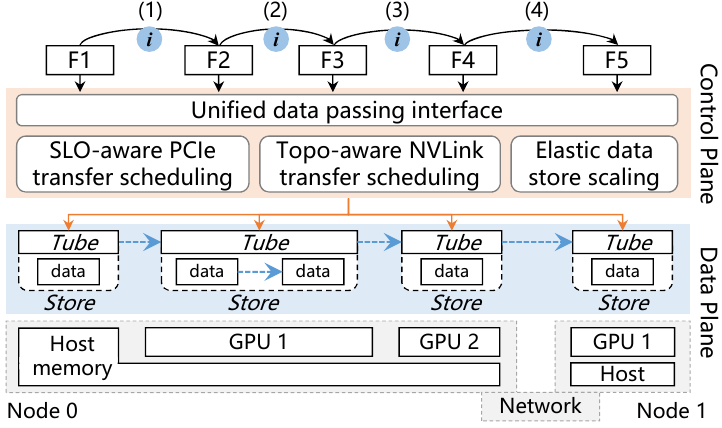}}
\caption{The overview of FaaSTube.}
\label{overview}

\end{figure}

We propose \emph{FaaSTube}, a GPU-oriented data transfer framework designed for serverless inference workflows. As shown in Fig.~\ref{overview}, FaaSTube acts as a transparent "tube" across GPUs, facilitating efficient data storage and transfer for functions.

Functions initiate data passing through a \emph{unified data passing interface (\textbf{Challenge \#4})}, which simplifies the management of various data passing in serverless inference workflows and determines the optimal data transfer methods for different scenarios. FaaSTube's control plane oversees transfer scheduling and storage management through three core components: First, \emph{SLO-aware PCIe transfer scheduling (\textbf{Challenge \#1})}, which partitions the bandwidth of global PCIe connections to ensure performance isolation among concurrent functions; Second, \emph{Topology-aware NVLink transfer scheduling (\textbf{Challenge \#2})}, which exploits parallel NVLink paths to facilitate robust inter-function data passing across different GPU topologies; and Third, \emph{Elastic data store (\textbf{Challenge \#3})}, which elastically manages GPU memory and migrates intermediate data to reduce GPU memory usage. The data plane executes the decisions of the control plane, with a daemon thread running on each GPU and host that acts as both a \emph{tube}, responsible for transferring data within GPU servers, and a \emph{Store}, responsible for managing GPU memory and intermediate data.

\begin{lstlisting}[language=C++, caption={The APIs of user library}, label={code_example}][t]
void FaaSTube.unique_id(char** data_index);
void FaaSTube.fetch(char** index, void* input);
void FaaSTube.store(char** index, void* output, int response=0);

\end{lstlisting}


\section{Unified Data Passing Interface}

First, FaaSTube offers a unified data passing interface (List.~\ref{code_example}) to simplify the composition of serverless inference workflows. Second, FaaSTube enhances scalability through a two-tier distributed design.

\subsection{Unified Programming Interface}
FaaSTube provides a unified data ID to simplify managing host-side storage (e.g., shared host memory, Redis servers) and GPU-side storage (e.g., CUDA IPC handles). FaaSTube maintains a mapping between data IDs and corresponding address information, allowing functions to transfer data by simply passing the data ID. When a function calls \emph{store()}, FaaSTube saves the data locally by default, such as in the current GPU's memory, and updates the mapping table. When a function calls \emph{fetch()}, FaaSTube locates the data in GPU servers using the data ID and selects an appropriate transfer method based on the locations of data and the requesting function. FaaSTube supports four transfer methods (Fig.~\ref{overview}):

\begin{itemize}[leftmargin=*]

\item \underline{\emph{Host-GPU}}: For data transfer between host and GPU memory, the data is partitioned according to the number of PCIe connections in the GPU server and transferred in parallel using a pipelined approach. Specifically, the data is divided into smaller chunks, and once the first chunk reaches the neighboring GPU, it is immediately transferred to the target GPU. Notably, PCIe bandwidth usage for each function is managed by the PCIe scheduling module.

\item \underline{\emph{Intra-GPU}}: When a function fetches data on the same GPU, it can be directly mapped and copied into the function's address space using CUDA IPC~\cite{cudaIPC}.

\item \underline{\emph{Inter-GPU}}: When a function fetches data from a different GPU within the same node, FaaSTube will select appropriate NVLink paths based on the GPU topology. For example, on a non-uniform GPU topology, FaaSTube will use multiple NVLink paths to enable parallel data transfer.

\item \underline{\emph{Inter-node}}: When a function fetches data across different nodes, the data is transferred over the network. This involves copying the data to host memory, transmitting it over the network to the target node's host memory, and subsequently copying it to the target GPU. Note that FaaSTube conducts these data copies in a pipelined manner, whereas the host-oriented approach conducts them sequentially. While high-end GPUs support direct GPU RDMA, many cost-effective GPU servers lack this feature. Further exploration of GPU-direct RDMA is left for future work.
\end{itemize}

\subsection{Scalable Distributed Framework}
To ensure system scalability, FaaSTube uses a two-level mapping table: a local table for each node and a global table maintained by a central node. Functions within each node first query and update their local table; if the data ID is not found, they then query the global table. This method minimizes cross-node messaging and reduces latency. To ensure consistency, local mapping tables are periodically synchronized with the global table. Additionally, FaaSTube and functions use a shared communication channel (i.e., Linux pipe) for fast interface invocation, further reducing latency.

\section{Efficient Transfer Scheduling }
We present how FaaSTube schedules inter-function data passing to fully utilize various connections in GPU servers.
\subsection{SLO-aware PCIe Transfer Scheduling }
To leverage global PCIe connections within GPU servers while avoiding contention, FaaSTube abstracts global PCIe bandwidth and regulates each function's usage similar to rate control in computer networking~\cite{TGS}, achieving dynamic PCIe bandwidth partition among functions. 

\begin{figure}[t]
\centerline{\includegraphics[width=0.48\textwidth]{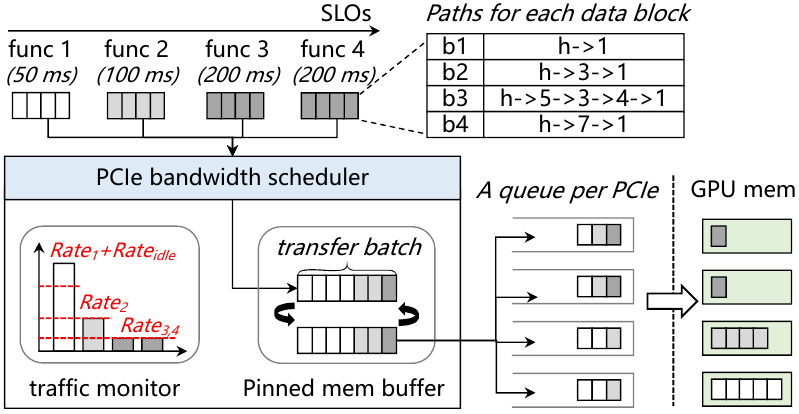}}
\caption{SLO-aware PCIe data transfer scheduling.}
\label{design2}
\end{figure}

Fig.~\ref{design2} shows the process of PCIe transfer scheduling in FaaSTube. First, each function's data is divided into smaller chunks (default size is 2MB) to enable fine-grained transfer control. FaaSTube allocates the PCIe bandwidth required by each function to meet its SLO requirements and triggers the transfer of data chunks from different functions in proportion. Finally, FaaSTube employs a circular pinned memory buffer shared among functions to improve transfer efficiency.






\noindent\textbf{Function transfer rate control.}
FaaSTube first calculates the minimum required transfer rate $Rate_{least}$ for each function based on its SLO and data size, representing the minimum bandwidth necessary to meet each function's SLO. Let $L_{slo}$ denote the function's SLO, and $L_{infer}$ denote its inference computation latency. The $Rate_{least}$ is defined as $data\_size/(L_{slo} - L_{infer})$. The predictability of DNN inference has been extensively studied~\cite{Astraea,GPUlet,streambox,Batch}. Given that FaaSTube utilizes temporal GPU sharing among functions, interference is minimized. Consequently, offline profiling can effectively guide transfer control to meet the latency SLOs for each functions.

FaaSTube monitors the transfer rate of each function's data block in real time to ensure it remains above $Rate_{least}$. FaaSTube then calculates the idle transfer rate (i.e., bandwidth) $Rate_{idle}$, which reflects the remaining bandwidth after meeting the minimum bandwidth requirements of all functions. Let $BW_{all}$ denote the total PCIe bandwidth in the GPU server, $Rate_{idle} = BW_{all}-\sum_{i=0}^{all\_funcs}{Rate_{least}^{i}}$. FaaSTube allocates this idle bandwidth to the function with the tightest SLO, enabling latency-sensitive functions to complete their data transfers first without impacting other functions. 




\noindent\textbf{Batched transfer triggering.} Since data chunk transfers cannot be interrupted once triggered, initiating all transfers simultaneously prevents newly arrived functions from preempting bandwidth. Conversely, triggering each chunk individually incurs additional overhead. Therefore, FaaSTube triggers data chunk transfers in batches (5 chunks by default), allowing newly arrived functions to preempt bandwidth by including their data chunks in subsequent batches.

\noindent\textbf{Circular pinned mem buffer.} To minimize the overhead of pinned memory allocation, a naive approach would cache all required pinned memory for each function; however, this requires substantial pre-allocation and may impair host system performance~\cite{pinnedmem}. Instead, FaaSTube employs a circular pinned memory buffer shared among functions, in conjunction with with batched data transfers, enabling different batches to reuse this fixed buffer and reducing the amount of cached pinned memory.


\begin{algorithm}[t]
  \caption{Contention-aware paths selection}\label{alg}
  \KwIn{Func\_id $func$; Source GPU $g_{s}$; Destination GPU $g_{d}$; The real-time global bandwidth usage matrix $BW_{nxn}$, The topology matrix $Topo_{nxn}$}
  \KwOut{The available parallel transfer paths $Paths$}
        \While{$path == null$}{
        $path \gets$ next\_shortest\_path($BW_{nxn}$, $g_{s}, g_{d}$)\;
        \If{all edges in $path$ is idle}{
            $Paths \gets path$\;
            Update($BW_{nxn}, path, func$)\;
        }
        \If{$BW_{out}(g_{s}) == 0 \cup BW_{in}(g_{d}) == 0 $}{
            break\;
        }
        
    }
    \If{$BW_{out}(g_{s}) \neq 0 \cap BW_{in}(g_{d}) \neq 0 $}{
    \While{$path == null$}{
        $path \gets$ next\_busy\_path($BW_{nxn}$, $g_{s}, g_{d}$)\;
        bandwidth\_balancing($path, func, BW_{nxn}$)\;
        $Paths \gets path$\;
        \If{$BW_{out}(g_{s}) == 0 \cup BW_{in}(g_{d})  ==  0 $}{
            break\;
        }
        
    }

    }

    \KwRet{$Paths$}\;


        

\end{algorithm}

\subsection{Topology-aware NVLink Transfer Scheduling}\label{topo-sche}
FaaSTube leverages parallel NVLink paths to enhance point-to-point data passing in non-uniform topologies. FaaSTube introduce a contention-aware path selection algorithm that optimizes NVLink usage while minimizing bandwidth contention from other functions (avoiding path duplication). 



Given a serverless workflow, FaaSTube first applies the placement strategy in MAPA~\cite{MAPA} to assign functions to GPUs and maximise NVLink connections between functions. After the function placement is determined,  FaaSTube first allocates direct connection paths between the GPUs included in the serverless workflow. If these direct NVLinks are already occupied by other functions, FaaSTube will force the other functions to release the path and re-plan other paths. Then, FaaSTube searches for available free NVLink paths for each inter-GPU data transfer in the serverless workflow, starting with the GPU pair with the least bandwidth. FaaSTube maintains a bandwidth usage matrix $BW(g, b)$, where $g$ represents GPUs and $b$ is the available bandwidth between them. FaaSTube continuously monitors and updates global bandwidth usage in real-time on this matrix $BW(g, b)$, which is used to guide path selection. 

As shown in Algorithm~\ref{alg}. the selection process involves: 1) FaaSTube first searches for free paths to avoid contention with other functions (lines 1-7). When a free $path$ is found, the bandwidth usage matrix $BW(g, b)$ is updated. The bandwidth occupied by the $path$ determined by the NVLink with the smallest bandwidth along the path, denoted as $b_{min}(path)$. Thus, the update to $BW(g, b)$ subtracts $b_{min}(path)$ from the free bandwidth of each GPU pair on the path. 2) If all free paths are exhausted and the outgoing bandwidth of $g_s$ and incoming bandwidth of $g_d$ are not saturated, FaaSTube searches busy paths to see if bandwidth can be balanced between the current function and the one occupying the path (lines 8-14). FaaSTube compares the total bandwidth used by the running function and the current function, and checks whether the running function can switch to another path. If it is available, the busy path is assigned to the current function. Because a GPU server usually has 4-8 GPUs, after using path pruning and other loop-free path search acceleration, the overhead of path selection is less than 10us in our experiments.

Parallel NVLink transfers use the pipelined method as in PCIe transfers, and batched data transfers to adapt to path changes. However, the difference is that the FaaSTube needs to adapt to the different NVlink bandwidths (24 or 48 GB/s) of each path, so the FaaSTube transmits data chunks proportional to each path's bandwidth.

\section{Elastic GPU Data Store}
To efficiently manage the GPU data store, FaaSTube 1) provides an auto-scaling memory pool on each GPU, which can elastically scale based on the actual demand of functions, and 2) intelligently migrates data between host and GPU memory based on request queues.

\subsection{Auto-scaling Memory Pool}\label{elastic_mempool}
As noted in Challenge \#3, fluctuating intermediate data sizes necessitate on-demand memory allocation in the GPU data store. However, temporary memory allocation introduces significant overhead in data passing, as native GPU allocations (e.g., cudaMalloc and cudaFree) incur millisecond-level delays. Our experiments show that temporarily allocating memory increases data passing latency by 19\% (Section~\ref{ablation_evel}). 

Memory pooling can mitigate this issue; however, existing methods lack elasticity, leading to excessive memory consumption in serverless environments. ML frameworks like PyTorch~\cite{pytorch} and TensorFlow~\cite{tensorflow} cache pre-allocated memory blocks for later reuse but do not actively release unused memory blocks. When function workloads and intermediate data sizes vary dynamically, this approach leads to up to 4X memory occupation than actual demand in our experiments. While PyTorch permits manual reclamation of memory pools, it reclaims all memory blocks, introducing overhead in future allocations. Recent work, GMlake~\cite{GMlake}, reduces fragmentation in the memory pool by using CUDA virtual memory and unified 2MB memory chunks, but it still lacks elastic memory reclamation. Moreover, costly IPC operations on each chunk in GMlake introduce large overhead in data passing (up to 45ms in our experiments).


\begin{figure}[t]
\centerline{\includegraphics[width=0.48\textwidth]{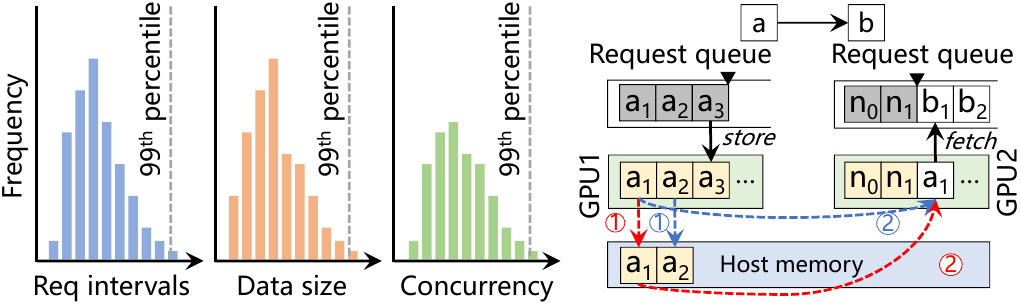}}
\caption{(a) Histogram policy characterizing both request arrivals (blue), intermediate data size (orange) and data accumulation (green) of each function. (b) Illustration of the inefficiency of LRU-based data migration (red line) vs. queue-aware data migration (blue line).}
\label{design3}

\end{figure}
FaaSTube introduces a new memory pre-warming strategy to dynamically rightsize the GPU memory pool for functions. Inspired by keep-alive strategies~\cite{infless,Azure-trace} used in function pre-warming, which track request intervals to define the duration each function stays in memory. However, the new challenge faced by FaaSTube is that the intermediate data size for the same function is fluctuating and may accumulate as function workload increases (\emph{Challenge \#3}). To accurately estimate memory requirements (Fig.\ref{design3}(a)), FaaSTube tracks not only request intervals ($R_{window} = Interval^{99th}$), but also the size of intermediate data  ($R_{size} = Data_size^{99th}$), and the degree of data accumulation ($R_{con} = Concurrency^{99th}$). After each function execution, memory reservation is calculated as $Data\_size=R_{size}\cdot R_{con}$. If no new requests arrive within the reservation window, the reserved memory is reclaimed. The total memory pool size is given by $MemPool\_size = \sum_{func} Data\_size\cdot1_{\{R_{window}\bigcap t\neq\varnothing \}}$, where $1_A$ is an index function of events, being 1 when event $A$ is true and being 0 otherwise. To accommodate bursty requests, FaaSTube maintains a minimum memory pool threshold, such as 300 MB in the experiment, instead of reducing it to 0.

\subsection{Smart Migration Guided by Request Queues}
Efficient data migration is crucial when intermediate data accumulates in GPU memory. Existing migration approaches, such as Unified Memory~\cite{Unifiedmem, HUVM, deepum}, rely on an LRU eviction strategy that removes earlier-stored (cold) data first. However, this method is ill-suited for managing intermediate data in serverless workflows, where earlier-stored data is often accessed sooner due to its downstream functions are enqueued earlier. This leads to unnecessary overhead from reloading data from host memory. For instance, as shown by the red line in Fig.\ref{design3}(b), the LRU strategy evicts the output data of function $a_1$ first, ignoring that $b_1$ (the downstream function of $a_1$) is enqueued earlier, forcing $b_1$ to reload data from host memory and introducing delays. 

Instead, FaaSTube adopts a queue-aware data migration approach, prioritizing the migration of intermediate data whose downstream functions are further back in the request queue. For example, as shown by the blue line in Fig.\ref{design3}(b), the output data of function $a_2$ is migrated before $a_1$'s output. Moreover, FaaSTube promptly clears intermediate data that is no longer needed and proactively reloads previously migrated data back to the GPU when sufficient memory becomes available. For instance, after $a_1$'s output is processed, $a_2$'s output is reloaded into GPU memory. In FaaSTube, data migrations are automatically triggered by memory pressure, such as reaching the data store's capacity limit (set to 1GB per GPU in our experiments), and are executed asynchronously with function execution.

\section{Implementation}
FaaSTube is built on INFless~\cite{infless}, a state-of-the-art serverless inference system. It comprises 5K lines of C++ code.

\noindent\textbf{Data transfer management}: Similar to Pheromone~\cite{pheromone}, FaaSTube mounts a shared host volume to each function for efficient data and message exchange on the host side. On the GPU side, FaaSTube launches a daemon thread for each GPU to handle data passing requests from functions. This thread uses CUDA IPC to establish a GPU memory mapping with the function, serving as a \emph{transfer buffer}. Since functions run in virtualized environments like containers~\cite{Nvidiacontainer}, the underlying GPUs are invisible to them. Data passing between functions is achieved via storing and fetching data through the transfer buffer. The daemon thread is responsible for transferring data between GPUs. The daemon thread manages data transfer between GPUs, utilizing multiple GPU streams to transfer data in parallel across different directions.

\noindent\textbf{Function scheduling}: FaaSTube follows the function scheduling in FaasFlow~\cite{faasflow} to minimize inter-node transfers in serverless workflow. Intra-node function placement follows existing strategy MAPA~\cite{MAPA}. Moreover, To mitigate performance impacts of function cold-starts, FaaSTube pre-warms required functions and models, as in SHEPHERD~\cite{SHEPHERD}.

\section{Evaluation}



\begin{figure}[t]
\centerline{\includegraphics[width=0.49\textwidth]{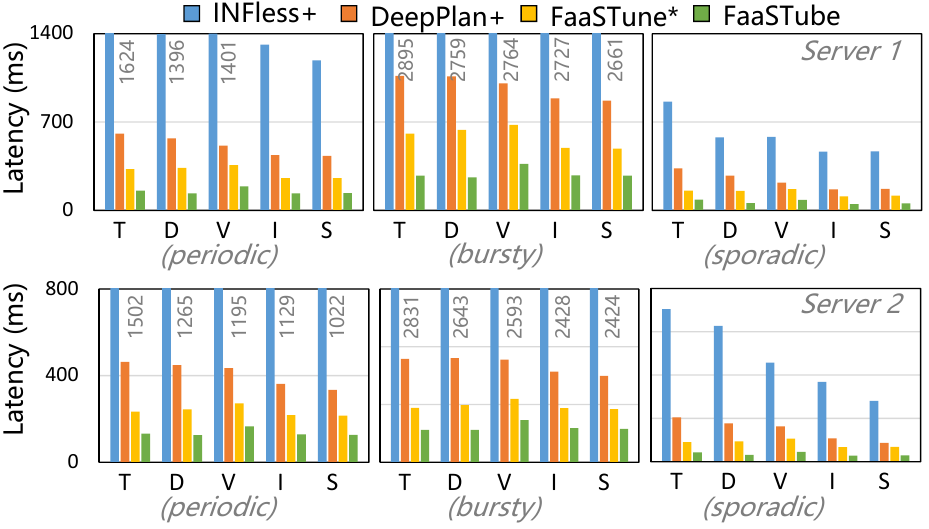}}
\caption{Comparison of the end-to-end latency on various workloads. Server 1 has 8xV100 and server 2 has 8xA100.}
\label{eval_end-to-end}
\end{figure}

\noindent\textbf{Setup.} Most experiments are conducted on two kinds of AWS GPU servers. \textbf{Server 1} (p3.16xlarge): This server has 8 NVIDIA V100 GPUs with NVLinks, a Xeon E5-2686 v4 CPU with 32 virtual CPUs, and 244GB of memory. \textbf{Server 2} (p4d.24xlarge): This server has 8 NVIDIA A100 GPUs with NVSwitch, a Xeon Plati. 8275CL CPU, and 1152GB of mem.

\noindent\textbf{Real-world inference workflows.} We conduct experiments using seven typical inference applications collected from the latest studies, as detailed below and in Table~\ref{workflows}. 
\begin{itemize}[leftmargin=*]\label{workflows_list}

\item \emph{Traffic} (T): Following Boggard~\cite{Boggard}, we implement a traffic monitoring workflow which first detects objects using the Yolo-det model, and then performs feature recognition on pedestrian and vehicle sub-images using ResNet models. 
\item \emph{Driving} (D): Following Adainf~\cite{Adainf}, we implement a road segmentation workflow for auto-driving. The process involves denoising the image, applying a semantic segmentation model, and outputting a colored image.
\item \emph{Video} (V): Following Aquatope~\cite{AQUATOPE}, we implement a video processing workflow that runs a face detection model on video chunks in parallel, followed by a recognition model to identify a specified actor.
\item \emph{Image} (I): Following Cocktail~\cite{cocktail}, we implement an image classification workflow that first denoises the image, then applies multiple classification models simultaneously, and aggregates the results to improve accuracy.
\item \emph{Social} (S): Following InferLine~\cite{infline}, we implement a social media workflow that identifies the text in user posts with an OCR model and analyzes it for sensitive content.
\item \emph{Yelp} (Y): Following Astraea~\cite{Astraea}, we implement a comment generation workflow that first analyzes comments for malicious content and then generates appropriate replies.
\end{itemize}

All pre-processing and post-processing are performed on the GPU using NVIDIA's CV-CUDA library~\cite{CVCUDA}. The datasets of these workflows are from~\cite{Adainf,Astraea}.

\noindent\textbf{Baselines.} We compare FaaSTube to the following baselines:
\begin{itemize}[leftmargin=*]

\item \emph{INFless+}: This baseline combines INFless~\cite{infless}, a state-of-the-art serverless inference system, with latest data passing optimization~\cite{pheromone,Fuyao} that accelerates data exchanges by enabling functions to share host memory.

\item \emph{DeepPlan+}: Derived from DeepPlan~\cite{deepplan}, this baseline uses parallel PCIe connections to accelerate data transfers in serverless workflows. Since DeepPlan focuses on model loading in monolithic inference systems, we incorporate its core ideas into INFless+. We refer to it as DeepPlan+.


\item \emph{FaaSTube*}: This baseline utilizes all available connections (NVLinks and parallel PCIe links) in GPU servers to accelerate data passing in serverless inference workflows, but lacks the further optimizations provided by FaaSTube. We refer to it as FaaSTube*.

\end{itemize}

\noindent\textbf{Workloads.}
We simulate the dynamic invocation of inference workflows, using production traces from Azure Function~\cite{Azure-trace}, which is a widely used trace in serverless research~\cite{Fuyao,infless,SHEPHERD}. There are three typical types of request arrival patterns in the production trace including sporadic, periodic, and bursty. As in Aquatope~\cite{AQUATOPE}, the load is scaled according to the resource availability (e.g., number of nodes and GPUs).

\begin{figure}[b]
\centerline{\includegraphics[width=0.48\textwidth]{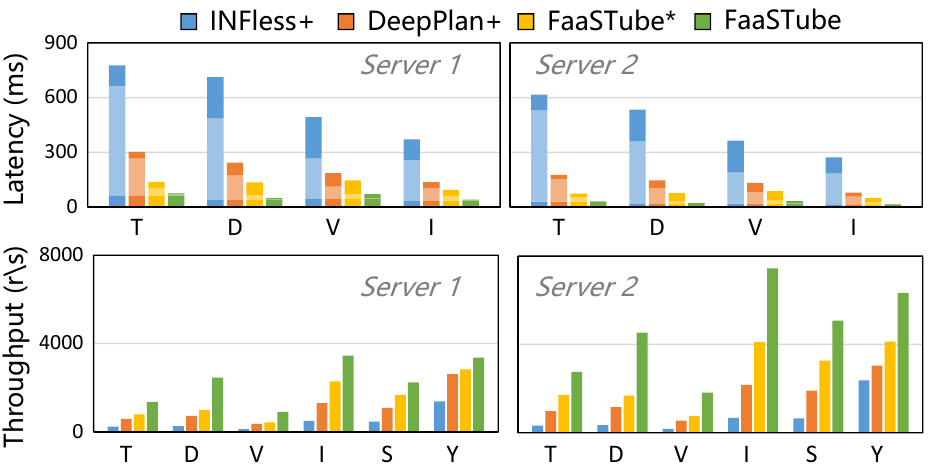}}
\caption{(a) Latency breakdown and (b) comparison of the maximum throughput.}
\label{eval_break_throughput}
\end{figure}

\subsection{Overall performance of FaaSTube}
In this section, we compare FaaSTube to state-of-the-art systems across a wide range of workflows.

\noindent\textbf{\uline{End-to-end latency}. FaaSTube can reduce the end-to-end latency of workflows by up to 90\%}.
Fig.~\ref{eval_end-to-end} shows the P99 latency across various workflows under different production workloads. FaaSTube consistently achieves the lowest latency, reducing it by 86\%-90\% compared to INFless+ and by 62\%-79\% compared to DeepPlan+. Both INFless+ and DeepPlan+ rely on host-oriented data passing, with DeepPlan+ using parallel PCIe connections, which still incur significant overhead due to redundant transfers between GPU and host memory. Furthermore, FaaSTube reduces latency by 43\%–63\% compared to FaaSTube*, which merely utilizes all available connections in GPU servers. This improvement is due to FaaSTube's optimized transfer scheduling and memory management during data passing.

\noindent\textbf{\uline{Latency breakdown}. FaaSTube can reduce the data passing overhead of workflows by up to 98\%}. To provide a detailed analysis of data passing overhead, we break down the P99 execution latency, excluding queuing time. As shown in Fig.~\ref{eval_break_throughput}(a), the latency is divided into three parts: host-to-gFunc data passing (top), gFunc-to-gFunc data passing (middle), and computation time (bottom). Results for bursty workloads are presented, as other workloads exhibit similar patterns. FaaSTube achieves the lowest data passing overhead, reducing it by 93\%–98\%, 90\%–94\%, and 70\%–88\% compared to INFless+, DeepPlan+, and FaaSTube*, respectively. DeepPlan+ leverages parallel PCIe data passing, resulting in lower latency than INFless+, which relies on a single PCIe link. FaaSTube* reduces gFunc-to-gFunc data passing overhead by utilizing NVLink. FaaSTube further reduces both types of data passing overhead through topology-aware transfer scheduling and efficient management of pinned memory and GPU memory during data passing.

\noindent\textbf{\uline{Throughput}. FaaSTube can improve throughput by up to 12X}. 
We measure the maximum throughput of these workflows. Fig.~\ref{eval_break_throughput}(b) shows that FaaSTube achieves the highest throughput, outperforming INFless+, DeepPlan+ and FaaSTube* by 2.4X–12X, 1.7X–3.9X and 1.3X–2.7X, respectively. The improvements are more pronounced in driving and video workflows, where large volumes of media data are transferred between functions and returned to host memory, making both gFunc-to-gFunc and host-to-gFunc data transfers significant bottlenecks. FaaSTube mitigates these bottlenecks by optimizing PCIe bandwidth utilization with circular pinned memory buffers and optimizing inter-GPU data transfers through topology-aware NVLink scheduling and elastic GPU storage. This shows that the throughput of serverless workflows is highly dependent on the efficiency of data transfers within GPU servers.




\subsection{Optimizations in FaaSTube}
Here, we show the effectiveness of each technique in FaaSTube.
\begin{figure}[b]
\centerline{\includegraphics[width=0.48\textwidth]{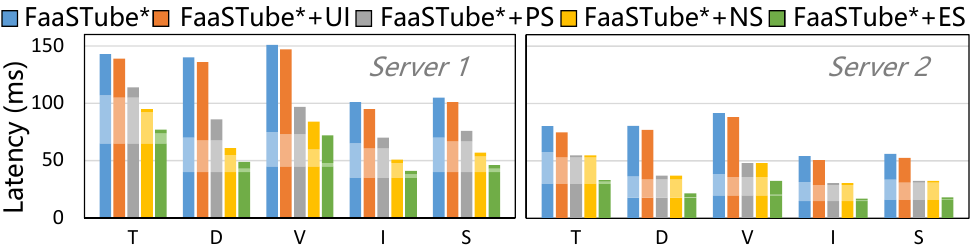}}
\caption{The latency when enabling each component in FaaSTube one by one.}
\label{eve_ablation}
\end{figure}

\subsubsection{Ablation Study.}\label{ablation_evel}
FaaSTube employs several techniques to enhance data passing efficiency: a unified data passing interface (UI), fine-grained PCIe scheduling (PS), topology-aware NVLink scheduling (NS), and an elastic data store (ES). We evaluate each technique by measuring the latency when enabling each one incrementally. As in the previous experiments, we report P99 latency under bursty loads, excluding queuing time.  As shown in Fig.~\ref{eve_ablation}(a), on the server 1, FaaSTube with all optimizations (rightmost bar) reduces latency by 46\%-65\% compared to FaaSTube*. Specifically, UI, PS, NS, and ES reduce latency by up to 2.5\%, 20\%, 23\%, and 19\%, respectively. UI's local messaging channel eliminates RPC overhead, while PS's circular pinned mem buffer reduces the cost of allocation in host-to-gFunc data passing. NS optimizes gFunc-to-gFunc data passing via parallel NVLink paths. Additionally, since host-to-gFunc data passing involves NVLink transfers with neighboring GPUs, NS helps reduce latency in those transfers as well. ES further reduces gFunc-to-gFunc data passing overhead by enabling rapid memory allocation through an auto-scaling memory pool and smartly migrating accumulated data to avoid the overhead of fetching from host (a 1GB limit is set for the GPU data store in the experiments). As shown in Fig.~\ref{eve_ablation}(b), on the server 2, FaaSTube achieves a latency reduction of 57\%-72\%. UI, PS, NS, and ES reduce latency by up to 3.2\%, 30\%, 0\%, and 39\%, respectively. The increased computing power of the A100 GPU makes data passing overhead more pronounced, particularly due to the high allocation costs of pinned memory and GPU memory. Additionally, the fully connected GPU topology among A100 GPUs limits the performance gains from NS.

\begin{figure}[t]
\centerline{\includegraphics[width=0.48\textwidth]{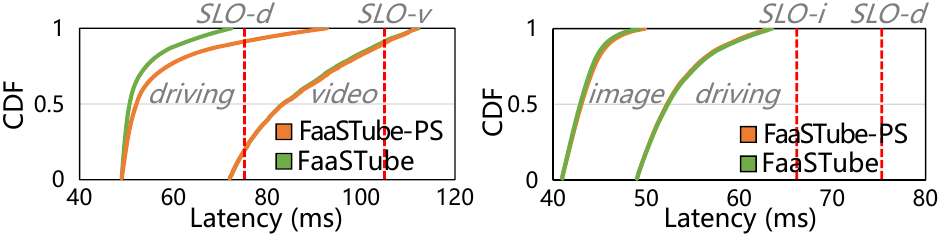}}
\caption{The effectiveness of FaaSTube’s PCIe transfer scheduling.}
\label{eval_pcieisolaton}
\end{figure}

\begin{figure}[t]
\centerline{\includegraphics[width=0.48\textwidth]{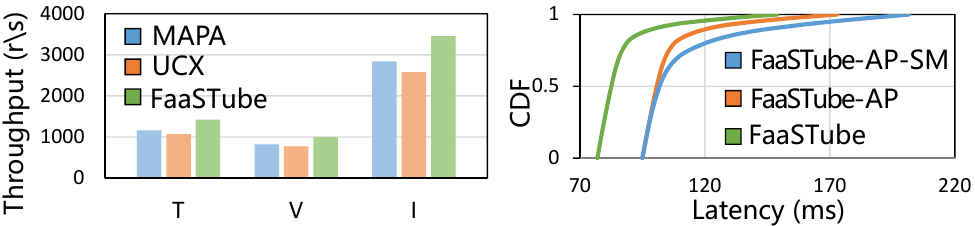}}
\caption{(a) The effectiveness of FaaSTube's parallel NVLink transfer and (b) The effectiveness of FaaSTube's elastic data store.}
\label{eval_nvlinkschedule_mempool}
\end{figure}

\subsubsection{PCIe Transfer Scheduling.} To demonstrate the effectiveness of FaaSTube's PCIe transfer scheduling (PS) in achieving performance isolation between functions, we conduct mixed workload experiments using two workflow pairs on server 1. The SLO for each workflow is set to 1.5X its independent runtime. We compare FaaSTube with FaaSTube-PS, which employs native PCIe bandwidth sharing as in DeepPlan+. Both workflows run under bursty workload, consistent with previous experiments. Fig.~\ref{eval_pcieisolaton}(a) presents the results for a high-contention case where the latency-critical driving workflow is paired with a transfer-intensive video workflow, which involves multiple functions loading video chunks simultaneously. Under native PCIe bandwidth sharing, driving's latency is increased due to interference from the video workflow. In contrast, FaaSTube reduces latency by 32\% and improves SLO compliance through fine-grained control of PCIe bandwidth usage for each function. Fig.~\ref{eval_pcieisolaton}(b) shows results for a low-contention scenario, where driving is paired with another real-time image workflow. In this case, FaaSTube and FaaSTube-PS performed identically, indicating that FaaSTube introduces no additional overhead in PCIe transfers.

\subsubsection{NVlink Transfer Scheduling.}
To demonstrate the advantage of FaaSTube's parallel NVLink transfer over methods like MAPA~\cite{MAPA}, which only optimize placement on non-uniform GPU topologies, we evaluate their maximum throughput on server 1. As shown in Fig.~\ref{eval_nvlinkschedule_mempool}(a), FaaSTube can improve the throughput of video, image and traffic workflows by 18\%, 13\%, and 17\% respectively compared to MAPA. We also compare a related work UCX~\cite{UCX}, which manually optimizes point-to-point transfers for HPC tasks on a 4xGPUs server. Due to the lack of topology-aware path selection in FaaSTube, its throughput is even lower than MAPA's.


\subsubsection{Elastic Data Store.}\label{experiments_memusage}


To demonstrate the advantage of FaaSTube's auto-scaling memory pool (AP) and smart migration guided by request queues (SM) in its elastic data store, we first analyze the impact of each component on latency. As shown in Fig.~\ref{eval_nvlinkschedule_mempool}(b), the auto-scaling memory pool reduces latency by an average of 19\%. For burst requests, smart data migration further reduces tail latency by 14\% by migrating non-immediately accessed data based on the request queue, avoiding the overhead of loading from the host memory.

\begin{figure}[b]
\centerline{\includegraphics[width=0.48\textwidth]{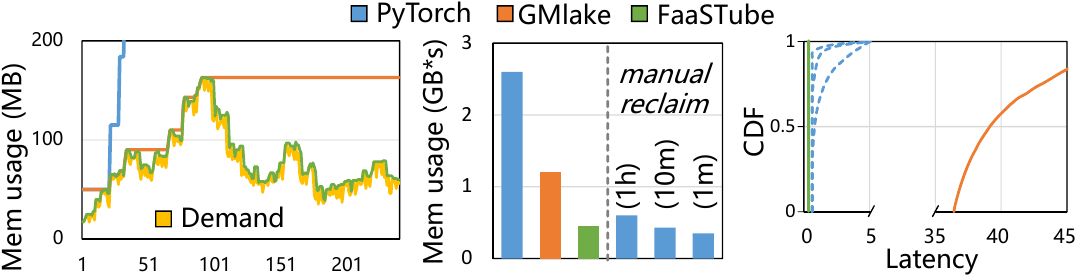}}
\caption{(a) Tracing the memory usage timeline, (b) Comparison of memory occupation and (c) Comparison of memory pooling efficiency. The blue dashed line refers to the performance of the pytorch memory pool under manual collection at different frequencies (1 hour, 10 mins and 1mins).}
\label{eval_memoryusage}
\end{figure}

Next, we compare FaaSTube with the latest pooling schemes in PyTorch~\cite{pytorch} and GMlake~\cite{GMlake}. As shown in Fig.~\ref{eval_memoryusage}(a-b), PyTorch caches all allocated memory blocks, leading to large memory occupation and fragmentation (e.g., a 100MB block cannot accommodate a 120MB request, requiring a new allocation) and does not actively free unused memory. GMlake mitigates fragmentation by using CUDA virtual memory and unified 2MB chunks but still lack active memory release, leading to large GPU memory consumption. In contrast, FaaSTube dynamically scales its memory pool based on function load and intermediate data size, thereby reducing memory usage.

While PyTorch allows manual memory pool reclamation, it reclaims all memory blocks, causing overhead in subsequent allocations. We evaluated memory allocation performance in PyTorch memory pool at varying reclamation frequencies (i.e., every hour, every ten minutes, and every minute). As shown in Fig.~\ref{eval_memoryusage}(b-c), while manual reclamation reduces memory usage, it increase tail latency by up to 4X (blue dashed curve). In contrast, FaaSTube dynamically scales the memory pool, effectively balancing memory usage and performance. The overhead of GMlake arises from the extra IPC operation costs introduced by each 2MB block during data sharing between the function and GPU data store.

\subsection{Scalability}
In this section, we evaluate the scalability and applicability of FaaSTube in a cluster and a GPU server without NVLink. 

\begin{figure}[t]
\vspace{0.25cm}
\centerline{\includegraphics[width=0.48\textwidth]{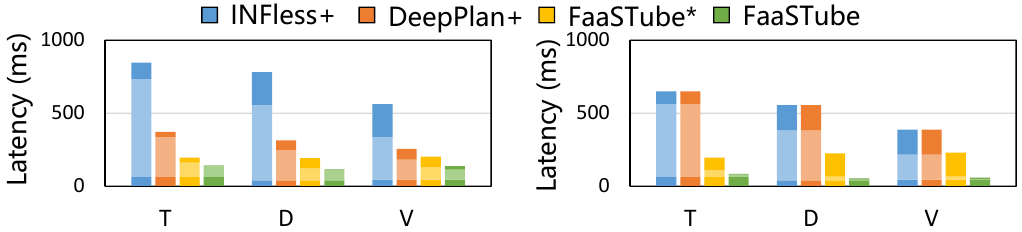}}
\caption{(a) Comparison of inter-node performance and (b) Comparison of performance on a 4xA10 GPUs server.}
\label{eval_scala}
\end{figure}

\noindent\textbf{Inter-node performance.}
Fig.~\ref{eval_scala}(a) shows the latency of deploying workflows on a cluster consisting of four AWS p3.16xlarge servers. As in the previous experiment, we report the P99 latency under bursty load, excluding queuing time. FaaSTube achieves optimal performance, reducing latency by 85\%, 63\%, and 39\% compared to INFless+, DeepPlan+, and FaaSTube* respectively. The improvement of FaaSTube comes from two parts: 1) intra-node transfers. Existing serverless workflow scheduling~\cite{faasflow} minimizes cross-node communication, resulting in at most one inter-node data transfer in a function workflow. Consequently, FaaSTube's PCIe and NVLink transfer scheduling remains effective for the remaining intra-node data transfers. 2) Inter-node transfers. INFless+ and DeepPlan+ utilize a host-oriented method, sequentially copying data between the source GPU, host storage on both nodes, and the target GPU, leading to significant cumulative latency. In contrast, FaaSTube maintains a comprehensive view of the cluster's connections, enabling rapid data transfer between GPUs across nodes using a pipelined approach, thereby reducing overhead. Additionally, FaaSTube's optimizations in pinned memory and GPU memory allocation further decrease inter-node data transfer overhead.



\noindent\textbf{Other GPU topologies.}
We evaluate FaaSTube's performance on a server with 4 A10 GPUs, which lack NVLink connections. Fig.~\ref{eval_scala}(b) shows that FaaSTube achieves the lowest latency, reducing it by up to 90\%, 90\%, and 75\% compared to INFless+, DeepPlan+, and FaaSTube*, respectively. On the A10 GPU server, each GPU can transfer data with host memory via only one single PCIe link, so DeepPlan+ and INFless+ perform the same. Additionally, INFless+ and DeepPlan+ are constrained by their host-oriented data passing methods, where data is sequentially copied to host storage and then to the target GPU, leading to latency accumulation. FaaSTube transfers data in a pipelined manner. FaaSTube* remains constrained by GPU memory and pinned memory allocation overhead during data transfers.

\section{Related Work}
\textbf{Multi-GPU communication}. Existing multi-GPU communication methods~\cite{nccl,TACCL,TCCL,MAPA,SCCL} primarily focus on collective communication primitives, such as allReduce, reduceScatter, and allGather, which use parallel rings or trees structures to integrate all GPUs. These methods overlook GPU point-to-point transfers on non-uniform topologies, which is needed by serverless inference workflows. For instance, NCCL's ncclSend/ncclRecv operations only use the direct NVLink path. Moreover, existing methods~\cite{MAPA,SCCL,Blink} for non-uniform topologies only addresses the problem partially through placement optimizations. In contrast, FaaSTube enables robust data transfers on non-uniform topologies by leveraging parallel NVLink paths with contention-aware path selection.

\textbf{GPU memory management.} Existing methods focus on pooling memory and unifying multi-level memory. Systems like GMlake~\cite{GMlake} use CUDA virtual memory to reduce fragmentation in memory pooling, while CUDA UVM~\cite{Unifiedmem}, HUVM~\cite{HUVM} and DeepUM~\cite{deepum} address GPU memory limits by swapping data between GPU and host memory. However, these methods are designed for long-running training tasks and lack flexible memory recycling, which can lead to large memory occupation in serverless context. In contrast, FaaSTube dynamically scales the memory pool according to the function's needs. Moreover, due to the limited visibility of underlying GPU resources, existing memory management techniques are not applicable to serverless functions.


\textbf{Serverless workflow optimizations.} Current research primarily focus on traditional workflows. Systems such as Pheromone~\cite{pheromone} and Unum~\cite{Domoreorless} optimize function composition, while Dataflower~\cite{dataflower} and Fuyao~\cite{Fuyao} improve data transfer in host memory, Nightcore~\cite{nightcore} minimizes runtime redundancy, and FaasFlow~\cite{faasflow} enhances function scheduling. Although these methods are orthogonal to FaaSTube, none address the need for efficient GPU-oriented data transfer in serverless inference workflows. In contrast, FaaSTube fully exploits various connections in GPU servers for serverless functions.

\textbf{GPU-enabled serverless system.} As demand for GPUs grows, many GPU-enabled serverless systems have emerged, typically categorized by how they share GPUs: temporal sharing (e.g., DGSF~\cite{DGSF} and FaaSwap~\cite{faaswap}) and spatial-sharing (e.g., StreamBox~\cite{streambox} and Llama~\cite{llama}). However, these systems do not optimize data transfers between GPU functions. Although FaaSTube follows a temporal-sharing model, its design can be extended to other GPU-enabled serverless systems.

\section{Conclusions}
We propose FaaSTube, an efficient GPU-oriented data passing framework for serverless inference workflows, enabling direct data exchange in GPU memory. FaaSTube optimizes transfer scheduling over PCIe and NVLinks to accelerate host-to-GPU and GPU-to-GPU data transfers among functions. Additionally, it implements an elastic memory pool on each GPU that automatically scales based on actual data passing demand of serverless functions, reducing GPU memory consumption.


\bibliographystyle{ACM-Reference-Format}
\bibliography{references}


\end{document}